\documentclass{PoS}

\usepackage{amsmath} 
\usepackage{float}
\usepackage{caption}
\usepackage{subcaption}
\usepackage{subfloat} 
\usepackage{slashed}
\newcommand{\gS}{\mathrm{S}}
\newcommand{\gP}{\mathrm{P}}
\newcommand{\gV}{\mathrm{V}}
\newcommand{\gA}{\mathrm{A}}

\newcommand{\sym}{\mathrm{sym}}
\newcommand{\gam}[1]{\mathrm{#1}}

\newcommand{\mlim}{\lim_{M_R \to \overline{m}}}
\newcommand{\dslash}{\not{\hbox{\kern-2pt $\partial$}}}
\newcommand{\pslash}{\not{\hbox{\kern-2pt $p$}}}
\newcommand{\kslash}{\not{\hbox{\kern-2pt $k$}}}
\newcommand{\qslash}{\not{\hbox{\kern-2pt $q$}}}

\newcommand{\Tr}{\mbox{Tr}}

\def\ie{{\it i.e.}\ }
\def\eg{{\it e.g.}\ }

\usepackage{lipsum}

\let\OLDthebibliography\thebibliography
\renewcommand\thebibliography[1]{
  \OLDthebibliography{#1}
  \setlength{\parskip}{0pt}
  \setlength{\itemsep}{0pt plus 0.3ex}
}

\title{A massive momentum-subtraction scheme}

\ShortTitle{A massive momentum-subtraction scheme}

\author{Peter Boyle, Luigi Del Debbio, \speaker{Ava Khamseh}\\
  School of Physics and Astronomy, University of Edinburgh\\
  EH9 3JZ, Edinburgh, United Kingdom\\
  E-mail: \email{a.khamseh@sms.ed.ac.uk}}

\abstract{We introduce a new massive renormalization scheme, denoted
  mSMOM, as a modification of the existing RI/SMOM schem. We use SMOM
  for defining renormalized fermion bilinears in QCD at non-vanishing
  fermion mass. This scheme has properties similar to those of the
  SMOM scheme, such as the use of non-exceptional symmetric momenta,
  while in contrast to SMOM, it defines the renormalized fields away
  from the chiral limit. Here we discuss some of the properties of
  mSMOM, and present non-perturbative arguments for deriving some
  renormalization constants. The results of a 1-loop calculation in
  dimensional regularization are briefly summarised to illustrate some
  properties of the scheme.}

\FullConference{The 34th International Symposium on Lattice Field Theory\\
  24 -30 July 2015\\
  University of Southampton, UK}

\begin{document}

\section{Introduction}
Lattice QCD simulations allow the determination of physical
quantities, such as meson masses and decay constants as well as
operator matrix elements, non-pertubatively. Matrix elements measured
on the lattice are bare quantities and have to be renormalized before
the continuum limit is taken. The renormalization scale $\mu$, is
often chosen such that
\begin{equation}
am \ll a\mu \ll \pi
\end{equation}
where $m$ is the mass of the quark and $a$ is the lattice spacing corresponding to cut-off $\pi/a$.\\
Heavy quarks such as charm are currently simulated in order to
investigate their non-perturbative dynamics. The mass of the quarks in
these simulations are often the same order as the lattice
cut-off. This can make it difficult to find a clear separation between
the fermion mass, the renormalization scale $\mu$, and the lattice
cut-off. Therefore, it would be interesting to introduce a mass
dependent renormalization scheme, with the renormalization conditions
being imposed at a finite renormalized mass $m\to\bar{m}$, while
having the renormalized WIs to hold implying scale independence of the
conserved currents.

\section{Massive Renormalization Conditions}
Following Ref.~\cite{Sturm:2009kb} we will consider continuum
Minkowski space, with symmetric non-exceptional momenta
\begin{equation}
p_2^2=p_3^2=q^2=-\mu^2,
\end{equation}
where $p_2$ and $p_3$ are incoming and outgoing off-shell momenta of the vertex, $q^\mu=p_2^\mu-p_3^\mu$ is the momentum out of the vertex and $\mu^2>0$ is the renormalization scale. 
\begin{figure}[!ht]
  \centering
  \includegraphics[scale=0.9]{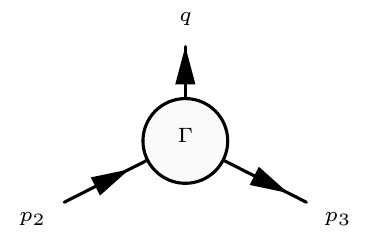}
  \caption{Kinematics used for the correlators of fermion bilinears.}
  \label{fig:kin}
\end{figure}
Vertex functions are defined as:
\begin{equation}
  \label{eq:Gcorr}
  G^a_\Gamma(p_3,p_2) = \langle O^a_\Gamma(q) \bar\psi(p_3) \psi(p_2)
  \rangle\, ,
\end{equation}
\ie as the correlator of two fermions together with the fermion
bilinear operator $O^a_\Gamma=\bar\psi \Gamma\tau^a\psi$, which is a
flavor non-singlet with $\tau^a$ denoting a generic generator of
rotations in flavor space. $\Gamma$ spans all the elements of the
basis of the Clifford algebra, which we denote as
$\Gamma=\gS,\gP,\gV,\gA,\gam{T}$. We consider a fermion doublet which
is degenerate in mass, $m_1=m_2=m$, and take
$\tau^a=\frac{\sigma^+}{2}=\frac12(\sigma^1+i\sigma^2)$ giving
$O_\Gamma=\overline{\psi}_1\Gamma{\psi_2}$. The amputated vertex
function is
\begin{equation}
  \label{eq:LambdaO}
  \Lambda^a_\Gamma(p_2,p_3) = S(p_3)^{-1} G^a_\Gamma(p_3,p_2) S(p_2)^{-1}\, ,
\end{equation}
where $S(p)$ is the fermion propagator:
\begin{equation}
  \label{eq:Sdef}
  S(p) = \frac{i}{\pslash - m - \Sigma(p) + i \epsilon}\, .
\end{equation}
Note that for each leg being amputated, the fermion propagator with
the corresponding flavor needs to be used.

Let us consider chiral symmetry transformations with a regulator that
does not break the symmetry, like \eg dimensional regularization. The
infinitesimal vector and axial non-singlet SU(2) transformations
change $\bar{\psi}$ and $\psi$ in the path integral, yielding the bare
vector and axial WIs. Written in terms of $\Lambda^a_\gV$ and
$\Lambda^a_\gA$ they are
\begin{align}
  \label{eq:VWIbare}
  q\cdot \Lambda^a_\gV &= i S(p_2)^{-1} - i S(p_3)^{-1} \, , \\
  \label{eq:AWIbare}
  q\cdot \Lambda^a_\gA &= 2 m i \Lambda^a_\gP - \gamma_5 iS(p_2)^{-1} - iS(p_3)^{-1} \gamma_5\, .
\end{align}

The renormalized quantities are defined as follows: 
\begin{equation}
  \label{eq:Zdef}
  \psi_R = Z_q^{1/2} \psi\, , \quad m_{qR} = Z_{m_q} m_q\, , \quad O_{\Gamma,R}
  = Z_\Gamma O_\Gamma\, ,
\end{equation}
yielding the renormalized propagator and amputated vertex functions 
\begin{equation}
  \label{eq:SGren}
  S_R(p) = Z_q S(p)\, , \quad \Lambda_{\Gamma,R}(p_2,p_3) =
  \frac{Z_\Gamma}{Z_q} \Lambda_\Gamma(p_2,p_3)\, ,
\end{equation}
where $q=l, H$ for light and heavy quarks respectively. Note that for each leg being amputated, the fermion propagator with the corresponding flavor needs to be used. In the rest of this section, we will denote $m_l$ by $m$ and ${m_H}$ by $M$ and suppress the flavor index $a$ to keep the notation simple.

The mSMOM renormalization condition are defined away from the chiral
limit, at some reference mass $\bar{m}$ which can be chosen freely:
\begin{align}
  \label{eq:mSMOM1}
  \mlim &\left. \frac{1}{12 p^2} \Tr \left[i S_R(p)^{-1}
     \pslash\right] \right|_{p^2=-\mu^2}=1\, ,\\
  \label{eq:mSMOM2}
  \mlim & \frac{1}{12 M_R} \left\{\left. \Tr \left[-i
     S_R(p)^{-1}\right] \right|_{p^2=-\mu^2}
     - \frac12 \left. \Tr \left[\left(q\cdot \Lambda_{\gA,R}\right) \gamma_5
     \right]\right|_\sym \right\}=1\, , \\
   \label{eq:mSMOM3}
  \mlim & \frac{1}{12 q^2} \Tr \left. \left[ \left(q
          \cdot \Lambda_{\gV,R} \right) \qslash \right] \right|_\sym =
          1\, , \\
   \label{eq:mSMOM4}
  \mlim & \frac{1}{12 q^2} \Tr \left. \left[ \left(q
          \cdot \Lambda_{\gA,R} - 2M_R i \Lambda_{\gP, R} \right)
          \gamma_5 \qslash \right] \right|_\sym = 1\, ,\\
   \label{eq:mSMOM5}
  \mlim &\frac{1}{12 i} \Tr \left. \left[ 
          \Lambda_{\gP,R} \gamma_5 \right] \right|_\sym = 1\, ,\\
   \label{eq:mSMOM6}
  \mlim &\Bigg\{\frac{1}{12}\text{Tr}\left[\Lambda_{S,R}\right]-
          \frac{1}{6q^2}\text{Tr}\left[2iM_R\Lambda_{P,R}\gamma_5\slashed{q}\right]\Bigg\}\Bigg|_{\text{sym}}=1\, .
\end{align}
As compared to the SMOM scheme~\cite{Sturm:2009kb}, only the
renormalization conditions for the axial and the scalar vertex
functions are modified, with the modifications being proportional to
the mass, and vanishing at tree-level. Therefore, in the chiral limit
$\bar{m}\to0$, these conditions reduce to SMOM exactly. Similar
to SMOM, the conditions for mSMOM are spelled out in such a way to
preserve the renormalized WIs and give $Z_V=Z_A=1$. The
non-perturbative derivation follows similar steps to those taken in
\cite{Sturm:2009kb}. For $Z_V$, using the relation between
renormalized and bare vertex functions, and Eq.~(\ref{eq:mSMOM3}), we
obtain
\begin{align}
  \mlim \frac{1}{12 q^2} &\Tr \left. \left[ \left(q
          \cdot \Lambda_{\gV} \right) \qslash \right] \right|_\sym =  
  \mlim \frac{Z_q}{Z_\gV} \frac{1}{12 q^2} \Tr \left. \left[ \left(q
          \cdot \Lambda_{\gV,R} \right) \qslash \right] \right|_\sym
  \\
  \label{eq:cfr1}
  & = \frac{Z_q}{Z_\gV}\, .
\end{align}
Using the vector Ward identity, Eq.~(\ref{eq:VWIbare}), the LHS of the
expression above can be written as
\begin{align}
  \mlim \frac{1}{12 q^2} &\Tr \left. \left[
  \left( i S(p_2)^{-1} - i S(p_3)^{-1}\right) \qslash 
  \right] \right|_\sym = 
  \frac{1}{12 q^2} \Tr \left. \left[
  i S(q)^{-1} \qslash 
  \right] \right|_\sym  \\
  \label{eq:cfr2}
  &= Z_q   \mlim \frac{1}{12 q^2} \Tr \left. \left[
  i S_R(q)^{-1} \qslash 
  \right] \right|_{q^2=-\mu^2} = Z_q  \, .
\end{align}
Comparing Eqs.~(\ref{eq:cfr1}) and~(\ref{eq:cfr2}) yields $Z_\gV=1$.

Because of the modified renormalization condition for the
renormalization of the axial vertex function, the computation of
$Z_\gA$ and $Z_M Z_\gP$ are coupled in the mSMOM scheme. The axial
Ward identity, Eq.~(\ref{eq:AWIbare}), can be rewritten in terms of
renormalized quantities:
\begin{align}
  \label{eq:RWI}
  \frac{1}{Z_\gA} q\cdot \Lambda_{\gA,R} &- \frac{1}{Z_M Z_\gP} 2 M_R
                                           i \Lambda_{\gP,R} = -
                                           \left\{ \gamma_5
                                           iS_R(p_2)^{-1} +
                                           iS_R(p_3)^{-1} \gamma_5 \right\}\, .
\end{align}
Two independent equations can be obtained by multipling
Eq.~(\ref{eq:RWI}) by $\gamma^5 \qslash$ and $\gamma_5$ respectively,
taking the trace, and evaluating correlators at the symmetric
point. In the first case we obtain
\begin{equation}
  \label{eq:ZAeq}
  (Z_\gA -1) = \left(1 - \frac{Z_\gA}{Z_M Z_\gP}\right) C_{mP}\, ,
\end{equation}
where
\begin{equation}
  \label{eq:mPdef}
  C_{m\gP} = \mlim \frac{1}{12 q^2} \Tr \left. \left[2i M_R
      \Lambda_{\gP,R} \gamma_5 \qslash\right]\right|_\sym\, .
\end{equation}
The second equation instead yields
\begin{equation}
  \label{eq:Adel}
  (Z_\gA - 1) C_{qA} = -2 Z_\gA \left(1 - \frac{1}{Z_M Z_\gP}\right)\, ,
\end{equation}
where we have introduced one more constant
\begin{equation}
  \label{eq:mA}Z_\gP
  C_{q\gA} = \mlim \frac{1}{12 M_R} \Tr \left. \left[
      q\cdot \Lambda_{\gA,R} \gamma_5
    \right]\right|_\sym \, .
\end{equation}
It is easy to verify that $Z_\gA=1$, $Z_M Z_\gP=1$ is the unique
solution of the system. It can be readily observed that given these
condition Eq.~(\ref{eq:RWI}) gives the correct renormalized axial
WI. In particular $Z_V$ and $Z_A$ are independent of the
renormalization scale $\mu$.

\section{Perturbative computation}
\label{sec:pert-comp}
We have checked all the above properties of the mSMOM scheme by
performing an explicit one-loop computation in perturbation theory
using dimensional regularization, keeping explicitly the dependence on
the bare mass $m$. We start from the 1-loop expression
\begin{equation}
  \label{eq:LgamOneLoop}
  \Lambda^{(1)}_\Gamma(p_2,p_3) = -i g^2 C_2(F) \int_k \frac{
    \gamma_\alpha \left[\pslash_3 - \kslash + m\right] \Gamma 
    \left[\pslash_2 - \kslash + m\right] \gamma^\alpha
  }{
    k^2\left[\left(p_3-k\right)^2-m^2\right] \left[\left(p_2-k\right)^2-m^2\right]
  }\, ,
\end{equation}
and rewrite the numerator in terms of scalar coefficients \cite{Chavez:2012kn, Smirnov:2008iw} multiplying
some form factors, depending on which vertex is being considered. The
scalar coefficients are functions of $m^2/\mu^2$. Computing the vertex
for all $\Gamma=S,P,V,A$ and the quark propagator at one loop, we
check that the results satisfy the bare WIs and reproduce the results in
Ref.\cite{Sturm:2009kb} as $m\to 0$. We then use the renormalization
conditions Eq.~\ref{eq:mSMOM1}-\ref{eq:mSMOM6}, and obtain
$Z_\gV = Z_\gA =1$, $Z_\gP=Z_\gS$, $Z_m Z_\gP=1$.

The details of this calculation goes beyond the scope of these
proceedings, and will be reported in details in a forthcoming
publication. 


\section{Mass non-degenerate scheme}
\label{sec:mass-non-degenerate-scheme}

We will now consider the renormalization scheme for the case of
non-singlet, mass non-degenerate vertex functions with the mass matrix
taking the form
\begin{equation}
  \label{eq:massmatrix-mixed}
  \mathcal{M} = \begin{pmatrix}
    M & 0\\
    0 & m
  \end{pmatrix} ,
\end{equation} 
where $M$ and $m$ are masses of the heavy and the light quarks
respectively. In what follows we will be interested in fermion
bilinears of the form $O^+=\overline{H} \Gamma l$ by choosing the
flavor rotation matrix to be
$\tau^a=\tau^+=\frac{\sigma^+}{2}=\frac12 (\sigma^1+i\sigma^2)$. For
clarity, we will leave the flavor index $``+"$ explicit in the WIs, but will suppress it for the rest of the section to keep
the notation simple. The curly letters
($\mathcal{V}, \mathcal{A}, \mathcal{P}, \mathcal{S}$) denote the
heavy-light bilinears. The vector and axial Ward identities are as
follows:
\begin{align}
  \label{eq:MixedVecWI}
  q \cdot \Lambda^+_{\mathcal{V}}=(M-m) \Lambda^+_{\mathcal{S}}+iS_{H}(p_2)^{-1}-iS_{l}(p_3)^{-1} \, , \\
  \label{eq:m=MixedAxialWI}
 q\cdot \Lambda^+_\mathcal{A} = (M+m) i \Lambda^+_\mathcal{P} - \gamma_5 iS_H(p_2)^{-1} - iS_l(p_3)^{-1} \gamma_5\, .
\end{align}

\subsection{Modified renormalization conditions}
\label{sec:modif-renorm-cond}

The mSMOM scheme for the heavy-light mixed case is defined by imposing
the following set of conditions at some reference mass $\overline{m}$:
\begin{align}
  \label{eq:mixedmSMOM3}
  \lim_{\substack{m_R\to0 \\ M_R\to\overline{m}}} & \frac{1}{12 q^2} \Tr \left. \left[ \left(q
                                               \cdot \Lambda_{\mathcal{V},R} -(M_R-m_R)\Lambda_{\mathcal{S},R} \right) \qslash \right] \right|_\sym =
                                               \lim_{\substack{m_R\to0 \\ M_R\to\overline{m}}}  \frac{1}{12 q^2} \Tr  \left[ \left(i\zeta^{-1}S_{H,R}(p_2)^{-1}-i\zeta S_{l,R}(p_3)^{-1}\right)
  \qslash \right] 
  \, , \\
  \label{eq:mixedmSMOM4}
  \lim_{\substack{m_R\to0 \\ M_R\to\overline{m}}}  & \frac{1}{12 q^2} \Tr \left. \left[ \left(q
                                                \cdot \Lambda_{\mathcal{A},R}- (M_R+m_R) i \Lambda_{\mathcal{P}, R}\right)
                                                \gamma_5 \qslash \right] \right|_\sym =
                                                \lim_{\substack{m_R\to0 \\ M_R\to\overline{m}}}    \frac{1}{12 q^2} \Tr  \left[ \left(
  -i\gamma^5\zeta^{-1}S_{H,R}(p_2)^{-1}-i\zeta S_{l,R}(p_3)^{-1}\gamma^5 \right)
  \gamma_5 \qslash \right] \, ,\\
  \label{eq:mixedmSMOM5}
  \lim_{\substack{m_R\to0 \\ M_R\to\overline{m}}}   &\frac{1}{12 i} \Tr \left. \left[ 
                                                 \Lambda_{\mathcal{P},R} \gamma_5 \right] \right|_\sym 
                                                 = \lim_{\substack{m_R\to0 \\ M_R\to\overline{m}}}  \Bigg\{
  \frac{1}{12(M_R+m_R)}\left\{\left. \Tr \left[-i
  \zeta^{-1} S_{H,R}(p)^{-1}\right] \right|_{p^2=-\mu^2}
  - \frac12 \left. \Tr \left[\left(q\cdot \Lambda_{\mathcal{A},R}\right) \gamma_5
  \right]\right|_\sym \right\}+  \nonumber  \\
                                             & \ \ \ \ \ \ \ \ \ \ \ \ \ \ \ \ \ \ \ \ \ \ \ \ \ \ \ \ \ \ \ \ \ \ \ \ \ \ \ \  \frac{1}{12(M_R+m_R)}\left\{\left. \Tr \left[-i
                                               \zeta S_{l,R}(p)^{-1}\right] \right|_{p^2=-\mu^2}
                                               - \frac12 \left. \Tr \left[\left(q\cdot \Lambda_{\mathcal{A},R}\right) \gamma_5
                                               \right]\right|_\sym \right\}
                                               \Bigg\}\, .
\end{align}
where $\zeta$ denotes the ratio of the light to the heavy field
renormalizations, i.e. $\zeta=\frac{\sqrt{Z_l}}{\sqrt{Z_H}}$. In the
degenerate mass limit $\zeta\to1$ and the mixed mSMOM prescription
reduces to the mSMOM and SMOM ones. The curly subscripts denote
heavy-light mixed vertices. The renormalization conditions for
$Z_l,\ Z_H$ and $Z_m$ remain unaltered as they are independently
determined from the corresponding degenerate, massive and
massless schemes of the previous sections. As usual the
renormalization conditions are satisfied by the tree level values of
the field correlators.

\subsection{Renormalization constants}
\label{sec:renorm-const}

The renormalization constants in this scheme are obtained once again
from the WIs. We multiply the vector WI
Eq.~\ref{eq:MixedVecWI} by $\qslash$, take the trace and write the
bare quantities in terms of the renormalized ones. Using
Eq.~\ref{eq:mixedmSMOM3} we obtain the solution $Z_{\mathcal{V}}=1$
and
\begin{align}
Z_\mathcal{S}=\frac{\frac{M_R}{Z_M}-\frac{m_R}{Z_m}}{M_R-m_R} \, .
\end{align}

For the axial current we follow a similar procedure, starting from the
bare mixed axial WI
Eq.~\ref{eq:m=MixedAxialWI}. Multiplying by $\gamma^5 \qslash$ and
$\gamma_5$ respectively and taking the trace gives two independent
equations. In the first case we use Eq.~\ref{eq:mixedmSMOM4} which
gives $Z_{\mathcal{A}}=1$ and
\begin{align}
\label{eqn:zpmixed}
Z_{\mathcal{P}}=\frac{\frac{M_R}{Z_MZ_\mathcal{P}}+\frac{m_R}{Z_mZ_\mathcal{P}}}{M_R+m_R}\, ,
\end{align}
as a solution. Note that in the degenerate mass limit, we recover
$Z_mZ_\gP=1$. In the second case, we take the trace with $\gamma^5$
and make use of Eq.~\ref{eq:mixedmSMOM5}, to obtain the solutions
$Z_{\mathcal{A}}=1$ and $Z_\mathcal{P}$ as in
Eq.~\ref{eqn:zpmixed}. One can easily check that this solution is
unique.

\section{Lattice regularization}
\label{sec:latt-regul}

If we use lattice as a regulator, the axial WI takes the form 
\begin{align}
  \label{eq:LatWardId}
  \nabla^*_\mu \langle A^a_\mu(x) \psi(y) \bar\psi(z) \rangle = 
  2 m & \langle P^a(x) \psi(y) \bar\psi(z) \rangle + \mathrm{contact\
  terms} \nonumber \\
  + \langle X^a(x) \psi(y) \bar\psi(z) \rangle\, .
\end{align}
where $X^a(x)$ is the explicit chiral symmetry breaking term due to
the lattice regulator. Since translational invariance of the action is
recovered in the naive $a\to 0$ limit, the contribution from the
explicit breaking by the regulator is given by by higher-dimensional
operators $X^a(x)=a O^a_5(x)$, where the suffix 5 indicates the fact
that these operators have classical dimension greater or equal to
5. In order to discuss the continuum limit of
Eq.~\ref{eq:LatWardId}, each term needs to be renormalized. In
particular the higher dimensional operator $O^a_5(x)$ mixes with the
lower dimensional ones appearing in Eq.~\ref{eq:LatWardId}:
\begin{align}
  \label{eq:O5Renorm}
  O^a_{5R}(x) = Z_5 \left[
  O^a_5(x) + \frac{\overline{m}}{a} P^a(x) + \frac{Z_\gA-1}{a} \nabla^*_\mu A^a_\mu(x)
  \right]\, .
\end{align}
According to Ref.~\cite{Testa:1998ez} the power divergences due to
this mixing do not contribute to the anomalous dimensions, \ie they do
not depend on the renormalization scale $\mu$ to all orders of
perturbation theory,
\begin{equation}
  \label{eq:AxRen}
  A^a_{R,\mu} = Z_\gA\left(g, am\right) A^a_\mu\, ,
\end{equation}
and the renormalized current satisfies the Ward identities up to terms
of order $a$. It is important to notice that the dependence of $Z_A$ on the
mass is a lattice artefact.

\section{Conclusion}
We have developed a massive renormalization scheme, mSMOM, for
non-singlet fermion bilinear operators in QCD with non-exceptional
momenta away from the chiral limit. The renormalization conditions are
imposed at some value $\bar{m}$ of the renormalized mass. In the limit
where $\bar{m}\to0$, our scheme reduces to the familiar SMOM
scheme~\cite{Sturm:2009kb}.

We have shown that the renormalized WIs for the case of both degenerate
and non-degenerate masses are satisfied non-perturbatively, giving
$Z_V = 1$ and $Z_A = 1$ for conserved currents. In order to gain a
better understanding of the properties of the mSMOM scheme we have
performed an explicit one-loop computation in perturbations theory
using dimensional regularization. Often on the lattice, we obtain
local currents that have to be renormalized. In this case the
renormalization constants can be obtained by taking ratios of vertex
functions with an appropriate projector and using both SMOM and mSMOM
conditions to extract $Z_V^\text{local}$ and $Z_A^\text{local}$. The
details of this procedure are deferred to a forthcoming publication. 

\acknowledgments{ We are indebted to Claude Duhr for his help with the technical aspects of massive one-loop computations and the use of his Mathematica package {\ttfamily PolyLogTools}.  AK is thankful to \linebreak Andries Waelkens and Einan Gardi for helpful discussions regarding the perturbative calculations. LDD is supported by STFC, grant ST/L000458/1, and the Royal Society, Wolfson Research
Merit Award, grant WM140078. AK is supported by SUPA Prize Studentship and Edinburgh Global Research Scholarship.  LDD and AK acknowledge the warm hospitality of the TH department at CERN, where part of this work has
been carried out. We are grateful to Andreas J\"uttner, Chris Sachrajda, Agostino Patella, Guido
Martinelli and Martin L\"uscher for comments on early versions of the manuscript. 
}

\end{document}